\documentclass[twoside,twocolumn,english,letterpaper,english,aps,prb,floatfix, showpacs, amsfonts, amssymb, superscriptaddress]{revtex4}
\usepackage[T1]{fontenc} 
\usepackage[latin1]{inputenc} 
\usepackage{amsmath} 
\usepackage{graphicx} 
\usepackage{amssymb} 
 
\makeatletter 

\usepackage{babel} 
\makeatother 
\begin{document} 
 
\title{Frustration of Decoherence in Open Quantum  Systems}

\author{E.~Novais}

\affiliation{Department of Physics, Boston University, Boston, MA 02215} 
 
\affiliation{Department of Physics, Duke University, Durham, NC, 27708}

\author{A.~H.~Castro~Neto}

\affiliation{Department of Physics, Boston University, Boston, MA 02215}

\author{L.~Borda}

\affiliation{Research Group of the Hungarian Academy of Sciences and Theoretical 
Physics Department, TU Budapest, H-1521, Hungary}

\affiliation{Sektion Physik and Center for Nanoscience, LMU M\"unchen, 80333 
M\"unchen, Theresienstr. 37, Germany}

\author{I.~Affleck}

\affiliation{Department of Physics and Astronomy, University of British Columbia, 
Vancouver, BC, Canada, V6T 1Z1} 
 
\author{G.~Zarand}

\affiliation{Research Group of the Hungarian Academy of Sciences and Theoretical 
Physics Department, TU Budapest, H-1521, Hungary}

\date{\today{}} 
 
\begin{abstract} 
We study a model of frustration of decoherence in an open quantum  system. 
Contrary to other dissipative ohmic impurity models, such as the Kondo model or 
the dissipative two-level system, the impurity  
model discussed here never presents 
overdamped dynamics even for strong coupling to the environment. We show that 
this unusual effect has its origins in the quantum mechanical nature 
of the coupling between the quantum impurity and the environment. 
We study the problem using analytic and numerical renormalization group 
methods and obtain expressions for the frequency and temperature  
dependence of the impurity susceptibility in different regimes. 
\end{abstract} 
 
 
\maketitle 
 
\section{introduction} 
 
In physics there is a large class of problems that  
can be described in terms of a single quantum mechanical degree of freedom  
interacting with an environment. 
Examples range from magnetic impurities in metals, superconductors, and 
magnets, macroscopic quantum tunneling in superconducting interference 
devices (SQUIDS) and molecular magnets \cite{mqt}, to qubits in quantum 
computers \cite{qc}. The common thread between all these  
problems is the dramatic effect that the dissipation has on the quantum  
dynamics of the impurity\cite{cl}. In particular, 
one of the most important effects of an environment on a quantum 
system is decoherence, that is, the destruction of quantum mechanical 
effects. Decoherence is the unavoidable 
consequence of the fact that no system in nature is really isolated. 
 
Impurity problems can be often 
reduced to an effective one-dimensional boundary problem that allows the use 
of powerful non-perturbative theoretical techniques. 
The Kondo model is 
probably one of the best known impurity problems and has been studied with  
a large number of theoretical tools, from the exact solution  
via Bethe ansatz \cite{bethe}, numerical renormalization group \cite{nrg},  
to conformal field theory \cite{cft}. The Kondo problem represents a  
universality class of open quantum systems where dissipation and 
decoherence play a fundamental role.  
In its anisotropic form, the Kondo effect can be mapped via dimensional  
reduction and abelian bosonization to the ohmic dissipative two-level system 
(DTLS) problem\cite{dtls}. The Kondo effect 
can be thought as a situation where decoherence is extreme, in the 
sense that the spin is completely screened by the environmental excitations 
in the formation of the so-called Kondo singlet. 
Moreover, impurities can be used as probes for the understanding of 
the environment itself and in some cases can even determine the properties 
of the environment in a self-consistent manner. This occurs in the case 
of the dynamical mean-field theories (DMFT) where the solution of a  
many-body problem 
reduces to the solution of a self-consistent impurity problem \cite{dmft}. 
Furthermore, systems where the competition between different phases 
of matter lead to the appearance of magnetic inhomogeneities  
(such as in the case 
of Griffiths-McCoy singularities in heavy fermion alloys) can many times be 
reduced to effective impurity problems \cite{griffiths}.  
  
In this paper we are going to describe a model for open quantum  systems 
that cannot be described within the Kondo universality class. This model 
describes an effect that we call frustration of decoherence where  
decoherence is reduced by a pure quantum mechanical effect. It is important, 
therefore, that one understands the physics behind the standard model 
of dissipation described by the Kondo or the DTLS and how it 
relates to the problem of decoherence. Since the connection between 
the Kondo problem and decoherence is not commonly discussed in the 
literature we will review some of the key features of the DTLS and 
its connection with the problem of decoherence. 
 
The DTLS can be described as a single spin half,  
${\textbf{S}}=(S_{1},S_{2},S_{3})$, 
coupled to a set of independent harmonic oscillators via the Hamiltonian 
 (we use units such that $\hbar=1=k_B$): 
\begin{eqnarray} 
H_{\mathtt{DTLS}} & = & \Delta S_{3} + \frac{i\lambda}{\sqrt{2 L}} 
S_{1} \sum_{k>0} \sqrt{k}(a_{k}-a_{k}^{\dagger}) \nonumber \\ 
& + & \sum_k vka_{k}^{\dagger}a_{k} , 
\label{eq:dtls} 
\end{eqnarray} 
where $\Delta$ is the tunnel splitting between the eigenvalues of $S_{1}$, 
$\lambda$ is the coupling to an environment of bosons with one-dimensional 
momentum $k$, and energy dispersion $\omega_{k}=vk$ ($v$ is the 
velocity of the excitations that we set to unity, $v=1$, from now on)  
and creation and annihilation operators 
$a_{k}^{\dagger}$ and $a_{k}$, respectively ($L$ is the linear 
size of the system). The operators obey canonical  
commutation relations:  
\begin{eqnarray} 
[a_{k},a_{k'}^{\dag}] &=& \delta_{k,k'} 
\nonumber 
\\ 
\left[ S_{i}, S_{j} \right] &=& i \epsilon_{ijk} S_{k} 
\label{comrel} 
\end{eqnarray}  
where $\epsilon_{ijk}$ is the Levi-Civita antisymmetric tensor.  
In this model one assumes 
a cut-off energy $\Lambda$, where $\Lambda$ is some non-universal 
quantity that is associated with microscopic properties of the bath 
($\Lambda$ is usually proportional to the inverse of the lattice spacing 
$a$).  
 
The physics described by Hamiltonian (\ref{eq:dtls}) can 
be summarize as follows. When $\vec{S}$ is decoupled from the 
environment ($\lambda=0$) one has an isolated spin problem in the 
presence of a {}``magnetic field'' proportional to $\Delta$. 
If at certain time $t=0$ the spin is prepared in an eigenstate of $S_{1}$, 
the {}``magnetic field'' induces transitions between 
the eigenstates of $S_1$ and the expectation value of the operator $S_1$, 
namely, $\langle S_1(t) \rangle$, oscillates harmonically with 
frequency $\Delta$. There is no release mechanism for the energy  
in the spin. By switching on a small coupling to the bath of oscillators,  
the harmonic oscillations of $\langle S_1(t) \rangle$ become underdamped due to the dissipation. 
Second order perturbation theory indicates that the behavior 
of the system depends on a dimensionless coupling $\alpha =  \lambda^{2}/8\pi$. 
For $\alpha <1/2$ there are two main effects \cite{saleur}: the slow modes 
of the bath, that cannot follow the motion of the spin, lead to damping 
and therefore to an exponential 
decay of $\langle S_1(t) \rangle$; the fast modes of the bath, that 
can follow the motion of the spin, lead to a new renormalized oscillation  
frequency $\Delta_R < \Delta$. For $\alpha >1/2$  there is a crossover to an  
overdamped regime where oscillations disappear (effectively $\Delta_R \to 0$) 
and only exponential decay occurs. Finally, at $\alpha=1$ 
there is a true quantum {}``phase transition'', where the the impurity 
spin becomes localized in one of the eigenstates of $S_{1}$. In the 
Kondo language the change from delocalized to localized  
is equivalent to a Kosterlitz-Thouless transition (KT) between the 
Kondo problem with ferromagnetic coupling (that has a triplet ground 
state) and the Kondo problem with antiferromagnetic coupling (with 
a singlet as ground state). 
 
One of the most illuminating ways to describe the KT transition is via 
a perturbative renormalization group (RG) calculation in leading order 
in $\Delta/\Lambda \ll 1$.  
The RG proceeds in two steps. In the first step  
one reduces the cut-off energy of the bosonic bath from $\Lambda$ 
to $\Lambda-d\Lambda$ by tracing out high energy degrees of freedom. 
In a second step the dimensionless coupling constants  
$\alpha$ and $h=\Delta/\Lambda$ 
are rescaled to the new cut-off leading to the RG equations\cite{dtls}:  
\begin{subequations} 
\label{kt} 
\begin{eqnarray} 
\frac{d\alpha}{d\ell} & = & -h^{2}\alpha \, , 
\label{kt1} 
\\ 
\frac{dh}{d\ell} & = & (1-\alpha)h \, 
\label{kt2} 
\end{eqnarray} 
\end{subequations} 
where $d\ell=d\Lambda/\Lambda$. Thus, for $\alpha>1$ the system scales 
under the RG to weak coupling ($h(\ell) \rightarrow 0$), and at 
low energies the tunneling splitting $\Delta(\ell)$ scales towards zero 
leading to localization. Conversely, for $\alpha<1$ the 
couplings scales towards strong coupling ($h \rightarrow \infty$) 
indicating that RG breaks down. 
The renormalization scheme fails at a certain energy scale 
(that is, the value of $\ell=\ell^{*}$ for which $h(\ell^{*}) 
\approx 1$). This characteristic scale is called the Kondo temperature 
that can be obtained directly from (\ref{kt}) as: 
$T_{K}\approx\Lambda(\Delta/\Lambda)^{1/(1-\alpha)}$. In the Kondo 
problem, for frequencies and temperatures below $T_K$ there is 
no reminiscence of the original impurity spin. This is an extreme 
example of decoherence. 
 
Although the RG equations clearly captures the asymptotic behavior 
of the spin dynamics, in order to observe the cross-over from underdamping  
to overdamping, one has to look at the frequency and temperature dependence  
of the spin correlation functions. This is even more important 
in the context of decoherence, since we are interested in measuring 
observables associated with the local degrees of freedom, not with 
the environment. In a spin problem, a particular apropos object is 
the impurity transverse susceptibility that is given by: 
\begin{eqnarray} 
\chi_{\perp}(\omega) & = & -i \int_0^{\infty}  
\frac{dt}{2 \pi} e^{i\omega t}\langle [S_{1}(t),S_{1}(0)]\rangle\,. 
\label{eq:xi-trans} 
\end{eqnarray} 
The imaginary part of $\chi(\omega)$, $\chi^{\prime\prime}(\omega)$, 
is a measure of the amount of energy that is dissipated from the spin 
into the environment. In the absence of coupling to the environment 
($\lambda=0$ in (\ref{eq:dtls})) we have  
$\chi^{``}(\omega) \propto \delta(\omega-\Delta)$ 
indicating the spin {}``oscillates'' freely with frequency $\Delta$. 
When $\lambda>0$ two different effects occur in the frequency behavior 
of $\chi^{\prime\prime}(\omega)/\omega$: (1) instead of a Dirac delta 
function one finds a broadened peak and  $\chi^{\prime\prime}(\omega)/\omega$ 
becomes finite at  
$\omega=0$, indicating that the oscillations become damped; 
(2) the maxima moves from $\Delta$ to a renormalized value $\Delta_{R}$ 
due to {}``dressing'' of the spin by fast environmental modes. 
In the DTLS, the value of $\chi^{``}(\omega)/\omega|_{\omega \to 0}$  
and its width  $\delta\omega$ are set by the $T_{K}$:  
$\chi^{``}(\omega)/\omega|_{\omega \to 0} \propto 1/T^2_{K}$ 
and $\delta\omega\propto T_{K}$. In particular, in the overdamped 
regime ($\alpha >1/2$) the peak in $\chi^{``}(\omega)$ at finite  
frequency vanishes  
completely leaving a smooth function centered around $\omega=0$ \cite{costi}. 
 
In this paper we are going to study a model that can be considered a 
generalization of the DTLS (\ref{eq:dtls}): 
\begin{eqnarray} 
H & = & \sum_{k>0}k \left(a_{k}^{\dagger}a_{k}+b_{k}^{\dagger}b_{k}\right)+ 
\Delta S_{3} 
\nonumber 
\\ 
 & + &\frac{i}{\sqrt{2 L}}\sum_{k>0} \sqrt{k} \left\{ 
 \lambda_{1}S_{1}(a_{k}-a_{k}^{\dagger}) \right. \nonumber\\ 
 & + & \left. 
 \lambda_{2}S_{2} 
 (b_{k}-b_{k}^{\dagger}) 
\right\} 
\label{qfm} 
\end{eqnarray} 
where there are two independent dissipative baths labeled by operators $a_k$ 
and $b_k$ with couplings $\lambda_1$ and $\lambda_2$. Notice that 
(\ref{qfm}) reduces to the DTLS, eq.~(\ref{eq:dtls}),  
when one of the couplings $\lambda_1$ or 
$\lambda_2$ vanishes.  
At first sight, the only apparent difference between  
(\ref{qfm}) and (\ref{eq:dtls}) 
is the existence of an additional 
bosonic bath coupled to a third spin component. Thus, naively one would  
expect an enhancement of decoherence in comparison with the DTLS since more 
heat baths are present.  
This naive argument fails to grasp that both baths are {}``competing'' 
with each other for the {}``ordering'' of the impurity. While the 
coupling $\lambda_1$ {}``tries'' to localize the spin in an eigenstate of  
$S_1$, the coupling $\lambda_2$ also {}``tries'' to localize the spin in an 
eigenstate of $S_2$. However, we see from (\ref{comrel}) that  
the operators $S_1$ and $S_2$ do not 
commute with each other and therefore one cannot find a common eigenstate  
for the spin to localize in. This purely quantum mechanical effect leads 
to a \emph{less decoherent} environment. We will show that when  
$\lambda_{1,2}=\lambda$ and $\Delta/\Lambda\ll1$ the spin dynamics  
is always in the underdamped regime, regardless of the bare value of  
the coupling constants. In our previous publication we called this state 
of affairs the {}``quantum frustration of decoherence'' \cite{prl}.  
 
The Hamiltonian (\ref{qfm}) was originally obtained in the study of an spin 
$1/2$ impurity embedded in an environment of large spin-S in $d=3$ dimensions 
\cite{prl}. The mapping between these two problems is given in 
appendix~\ref{sec:Impurity-spin-in}. The magnetic environment 
has two effects in the dynamics of the impurity. The molecular fields 
produced by the environmental spins favor the alignment of the impurity spin 
along the ordering direction giving rise to a {}``magnetic field'' proportional 
to $\Delta$. The transverse magnetic fluctuations (spin waves) 
produce quantum fluctuations that tend to misalign the impurity spin leading 
to couplings proportional to $\lambda_{1,2}$ and therefore to dissipation.  
In an ordered antiferromagnetic spin environment the low energy,  
long-wavelength excitations, are two massless Goldstone modes  
(two transverse magnon 
excitations) that couple to the two different components of the spin 
as in (\ref{qfm}). The problem of impurities in magnetic media, especially 
in the paramagnetic phase, has received a lot of attention in the 
context of quantum phase transitions \cite{subir,vojta}. As we are going to 
show in what follows, the effect of quantum frustration occurs at finite 
energies or frequencies and therefore before the asymptotic regime is reached 
(very low frequencies) and the impurity spin fully aligns with the  
environmental 
spins. Thus, as in the case of the Kondo problem, quantum frustration is 
a crossover phenomenon that cannot be obtained in {}``asymptopia''.  
We should stress, however, that the phenomenon of quantum frustration is 
more general than its origin would imply. As in the case of the Kondo  
effect, it represents a universality class of impurity problems  
where decoherence is reduced by pure quantum mechanical effects. 
   
As mentioned above, impurity problems can be treated by powerful theoretical 
techniques when reduced to one-dimensional models with a boundary. 
It is convenient, therefore, to rewrite (\ref{qfm}) in a real space 
representation: 
\begin{eqnarray} 
H & = & \int_{-\infty}^{+\infty} 
dx\sum_{a=1,2}\left(\partial_{x}\phi_{a}(x)\right)^{2}+ 
\Delta S_{3}\nonumber \\ 
 & - & 
 \sqrt{8\pi\alpha_{1}}\partial_{x}\phi_{1}(0)S_{1}-\sqrt{8\pi\alpha_{2}}\partial_{x}\phi_{2}(0)S_{2}, 
\label{eq:hamil-0} 
\end{eqnarray} 
where $\phi_{1,2}(x)$ are one dimensional chiral bosonic fields (that is, 
left movers only)  
associated with the bosonic modes $a_k$ ($b_k$) and we have defined 
$\alpha_{1,2}=\lambda_{1,2}^{2}/8\pi$. 
We are ultimately interested in the general problem of decoherence described 
by (\ref{qfm}) or (\ref{eq:hamil-0}) and the mechanism of quantum frustration 
associated with this model. 
 
The paper is organized as follows: we derive the main RG 
equations in Section~\ref{sec:perturbative-renormalization-group} and 
show that the dissipative model discussed here is always coherent 
and shows scaling at strong coupling;  
in Section~\ref{sec:susceptibility-at-$T=3D0$} 
we study the impurity susceptibility using numerical renormalization group 
and analytical RG via the Callan-Symansky equations;  
Section~\ref{conclusions} contains a discussion of the problem of 
frustration of decoherence and also our conclusions. There are various 
appendices where the details of the calculations have been included. 
 
\section{renormalization group\label{sec:perturbative-renormalization-group}} 
 
Notice that, according to the RG equations (\ref{kt}), the KT transition 
occurs at a finite value of the coupling constant $\alpha$ and therefore cannot 
be obtained directly from perturbation theory. Instead, one has to use a 
rotated basis of states, obtained from a unitary trasnformation,  
where the problem becomes perturbative. This can be accomplished in 
our case by defining two unitary transformations: 
\begin{subequations} 
\label{rotation} 
\begin{eqnarray} 
U_{1} & = & 
e^{i\frac{\pi}{2}S_{2}}e^{i\sqrt{8\pi\alpha_1}\phi_{1}\left(x=0\right)S_{3,}} 
\label{eq:rotation-1}\\ 
U_{2} & = & 
e^{i\frac{\pi}{2}S_{1}}e^{i\sqrt{8\pi\alpha_2}\phi_{2}\left(x=0\right)S_{3}} 
\, , 
\label{eq:rotarion2} 
\end{eqnarray} 
\end{subequations} 
that rotate the impurity spin around the $S_3$ direction by angles that 
depend on the field configurations and around $S_2$ ($S_1$) by $\pi/2$. 
Notice that $U_1$ ($U_2$) generates a non-perturbative rotation in terms of 
the coupling $\alpha_1$ ($\alpha_2$).  
 
Let us consider the problem after rotation by $U_1$.  
By applying $U_{1}$ to the Hamiltonian (\ref{eq:hamil-0}), we obtain 
\begin{eqnarray} 
U_{1}^{\dagger}HU_{1} =  
H_{0} +\frac{1}{2}\left(\Delta A_{1}^{+}+i\sqrt{8\pi\alpha_{2}}B_{1}^{+}\right)+\mathtt{h}.\mathtt{c}. 
\label{eq:hamilkappa1} 
\end{eqnarray} 
where $H_0$ is the free bosonic Hamiltonian (the first term in the left hand 
side of (\ref{eq:hamil-0})). We have defined two vertex operators, 
\begin{subequations} 
\label{vertices} 
\begin{eqnarray} 
A_{1}^{\pm} & = & e^{\mp 
  i\sqrt{8\pi\alpha_{1}}\phi_{1}\left(x=0\right)}S^{\pm}, 
\label{vertices1} 
\\ 
B_{1}^{\pm} & = & \partial_{x}\phi_{2}\left(x=0\right)e^{\mp 
  i\sqrt{8\pi\alpha_{1}}\phi_{1}\left(x=0\right)}S^{\pm}, 
\label{vertices2} 
\end{eqnarray} 
\end{subequations} 
where 
$S^{\pm} = S_1 \pm i S_2$ are the standard raising (lowering) 
operators. 
 
As in the case of a generalized Coulomb gas problem \cite{cg,ayh},  
the partition function of the problem, $Z$, can be obtained in the basis 
that diagonalizes $S_3$ ($S_{3} |s_{3}\rangle = \pm \frac{1}{2} |s_3 \rangle$) as  
\begin{eqnarray} 
Z & = & \sum_{\left\{S^{z}\right\} }\int D\phi_{1,2}(x,\tau)\, 
e^{-S_{0}[\phi_{1,2}(x,\tau)]}\,\prod_{j}\frac{\delta\tau}{2}\left[ 
\Delta A_{1}^{m_{j}}\left(\tau_{j} 
\right)\right. 
\nonumber  
\\ 
 & + & 
 \left.i m_j \sqrt{8 \pi \alpha_{2}}B_{1}^{m_{j}}\left(\tau_{j}\right)\right] 
\label{eq:actionkappa1} 
\end{eqnarray} 
where $S_0$ is the action for the free boson fields, $\delta \tau$ is 
the time step in the imaginary time direction, and  
$m_{j}=s_{3}\left(\tau_{j}+\delta\tau\right)-s_{3}\left(\tau_{j}\right)$ 
is either $+1$ for a kink or $-1$ for an anti-kink at time $\tau_{j}$ 
of a given spin history in imaginary time. The partition function 
given in (\ref{eq:actionkappa1}) is the starting point of the RG analysis. 
 
We can define the Fourier transforms of the vertex operators, 
$A_1(\omega) = \int d\tau \exp\{i\omega \tau\} A_1(\tau)$ 
and bosonic fields $\phi_{1,2}(k,\omega) = \int dx \int d\tau 
\phi_{1,2}(x,\tau) \exp\{i(kx-\omega\tau)\}$,  
and divide the fields into slow modes, say $A_{1,<}(\tau)$,  
with $(\omega,k) < \Lambda$ and 
fast modes, say $A_{1,>}(\tau)$  
with $(\omega,k)>\Lambda$. We then integrate the fast modes 
within a shell $\Lambda<\left(k,\omega\right)<\Lambda+d\Lambda$, 
to obtain the renormalization of the slow fields due to the fast modes. 
In this procedure the renormalization of the slow modes is given by 
averages over the fast modes. It is straightforward to show that:  
\begin{subequations} 
\label{rgvertex} 
\begin{eqnarray} 
\left\langle A_{1}^{\pm}\left(\tau\right)\right\rangle _{>} & = & 
A_{1,<}^{\pm}\left(\tau\right)e^{- \alpha_{1}d\ell} 
\label{rgvertex1} 
\\ 
\left\langle B_{1}^{\pm}\left(\tau\right)\right\rangle _{>} & = & 
B_{1,<}^{\pm}\left(\tau\right)e^{-\left(1+  \alpha_{1}\right)d\ell} 
\label{rgvertex2} 
\end{eqnarray} 
\end{subequations} 
where, $d\Lambda/\Lambda=d\ell$ and $\langle P \rangle_>$ 
indicates the average of the operator $P$ over the fast modes. 
Substituting (\ref{rgvertex}) into (\ref{eq:actionkappa1}) and 
rescaling the fields in order to obtain the same partition 
function with slow modes only, we find that the couplings 
have to change with $\ell$ according to (see Appendix \ref{rgeq}): 
\begin{subequations} 
\label{eq:rg-1} 
\begin{eqnarray} 
\frac{d\alpha_{2}}{d\ell} & = & -2\alpha_{1}\alpha_{2} 
\label{eq:rg-1-k2} 
\\ 
\frac{dh}{d\ell} & = & \left(1-\alpha_{1}\right)h \,, 
\label{eq:rg-1-k3} 
\end{eqnarray} 
\end{subequations} 
which define the RG equations for $\alpha_2$ and $h$ but 
not for $\alpha_1$. The RG equation for $\alpha_1$ is obtained 
in second order in $h$.  
In the language defined by Anderson-Yuval-Hamann \cite{ayh}, 
it corresponds to the renormalization in $\alpha_{1}$ due to a {}``close 
pair'' of flip and anti-flip that is removed from a spin history 
in a particular RG step. One can show 
that a new operator, which is not present in the original problem is 
generated under this procedure\cite{Novais02}. This operator reads: 
\begin{eqnarray} 
C  \approxeq  
1-i\sqrt{2\pi\alpha_{1}}h^{2}\partial_{\tau}\phi_{1}\left(0,\tau\right)S_{3} 
\left(\tau\right)d\ell\delta\tau. 
\end{eqnarray} 
This term can be reexponentiating into the action, Eq.~(\ref{eq:actionkappa1}), 
and then integrated by parts in $\tau$. The final result is equivalent to a  
redefinition of the vertex operators, 
\begin{subequations} 
\label{newvertex} 
\begin{eqnarray} 
A_{1}^{\pm} & = & e^{\mp 
  i\sqrt{8\pi\alpha_{1}}\left(1-\frac{1}{2}h^{2}d\ell\right)\phi_{1}\left(0\right)}S^{\pm}, 
\\ 
B_{1}^{\pm} & = & 
\partial_{x}\phi_{2}\left(0\right)e^{\mp 
  i\sqrt{8\pi\alpha_{1}}\left(1-\frac{1}{2}h^{2}d\ell\right)\phi_{1}\left(0\right)}S^{\pm}, 
\end{eqnarray} 
\end{subequations} 
immediately implying the RG equation for $\alpha_{1}$ \cite{note_ope},  
\begin{eqnarray} 
\frac{d\alpha_{1}}{d\ell} & = & -h^{2}\alpha_{1}. 
\label{eq:rg-1-k1} 
\end{eqnarray} 
 
\begin{figure}[htbp] 
\includegraphics[%
  width=0.80\columnwidth]{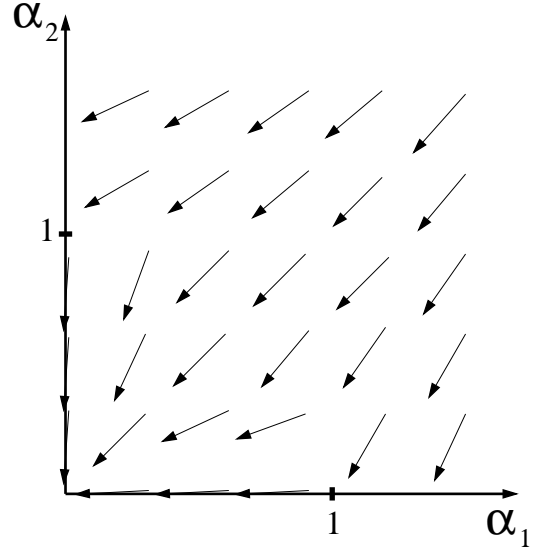} 
\caption{\label{cap:RGflow-1}Renormalization group flow given by 
  eqs.~(\ref{finalrg}) in the $\alpha_1$ versus $\alpha_2$ plane.} 
\end{figure}

Eqs.~(\ref{eq:rg-1}-\ref{eq:rg-1-k1}) where derived by a perturbative 
treatment in powers of $\alpha_{2}$ and $h$ and are valid up to second 
order in these coupling with $\alpha_1$ being arbitrary.  
If instead we apply the unitary transformation 
Eq.~(\ref{eq:rotarion2}) a similar set of equations can be derived for  
$\alpha_{1}$ and $h$ small with $\alpha_2$ being arbitrary. Notice that 
the only change in the RG equations is the interchange between  
$\alpha_1$ and $\alpha_2$ in  
(\ref{eq:rg-1-k2}-\ref{eq:rg-1-k1}). In fact, given the form of 
the the Hamiltonian (\ref{qfm}) it is easy to see that the RG 
equations must be symmetric under the interchange of $\alpha_1$ 
and $\alpha_2$. Thus, it is straightforward to see that by 
symmetry the RG equations are: 
\begin{subequations} 
\label{finalrg} 
\begin{eqnarray} 
\frac{d\alpha_{1}}{d\ell} & = & 
-2\alpha_{1}\alpha_{2}-\alpha_{1}h^{2}, 
\label{finalrg1} 
\\ 
\frac{d\alpha_{2}}{d\ell} & = & 
-2\alpha_{2}\alpha_{1}-\alpha_{2}h^{2}, 
\label{finalrg2} 
\\ 
\frac{dh}{d\ell} & = & (1-\alpha_{1}-\alpha_{2})h. 
\label{finalrg3} 
\end{eqnarray} 
\end{subequations} 
The symmetrization process is just a simple way to obtain the next 
order corrections to the RG equations. Strictly speaking, the RG 
equations (\ref{finalrg}) are valid up to second order in $h$,  
when either both $\alpha_1$ and $\alpha_2$ are of the same order and  
small, or when one of them small and the other is arbitrary. However, the  
terms of the form 
$\alpha_{1}\alpha_{2}$ could also be directly obtained 
from a diagrammatic technique \cite{zarand}. Notice that in the 
highly anisotropic case, say $\alpha_2 = 0$ ($\alpha_1=0$),  
we identify $\alpha_1 = \alpha$ ($\alpha_2=\alpha$)  
so that eq.~(\ref{finalrg1}) (eq.~(\ref{finalrg2})) 
reduces to eq.~(\ref{kt1}) and eq.~(\ref{finalrg3}) becomes 
(\ref{kt2}). As expected, our problem maps into the DTLS and one 
obtains a KT transition at $\alpha_1 = 1$ ($\alpha_2=1$).  
The RG flow associated with eqs.~(\ref{finalrg}) in the $\alpha_1$ 
versus $\alpha_2$ plane for fixed $h$ is shown in  
Fig.~\ref{cap:RGflow-1}. 
 
In the fully symmetric case where $\alpha_{1}=\alpha_{2}=\alpha$ one 
finds a very different physics. Indeed, from (\ref{finalrg}), one 
gets: 
\begin{subequations} 
\label{rgiso} 
\begin{eqnarray} 
\frac{d\alpha}{d\ell} & = & -2\alpha^{2}-\alpha h^{2}\,, 
\label{rgiso1} 
\\ 
\frac{dh}{d\ell} & = & (1-2\alpha) h \, . 
\label{rgiso2} 
\end{eqnarray} 
\end{subequations} 
As one can see from Fig.~\ref{cap:RGflow-1} there is no KT transition 
in this case. The couplings $\alpha_1$ and $\alpha_2$ always flow 
to zero while $h$ scales towards strong coupling. In the DTLS 
language the spin never localizes in an eigenstate of $S_1$ or $S_2$ 
being always in an eigenstate of $S_3$. Hence, in the isotropic case, 
no matter how large the couplings to the environment the spin is 
always coherent. This is the phenomenon of quantum frustration of decoherence. 
 
We can obtain a more quantitative analysis of the RG scale in 
some particular limits.  
As noticed before the RG breaks down at a scale $\ell^* = \ln(\Lambda_0/T_A)$ 
(where $\Lambda_0$ is the initial cut-off of the problem) 
when $h(\ell^*) \approx 1$. $T_A$ is the crossover energy  
scale from weak to strong coupling (the equivalent of the Kondo temperature). 
It is easy to see that the value of $T_A$ depends on the bare value 
of $\alpha(\ell=0)$. If $\alpha(0) \ll h(0)$ the flow is  
essentially the same as the usual KT flow and one can disregard the 
flow of $\alpha(\ell)$ in order to find,  
\begin{eqnarray} 
T_A \approx  \Delta \left({\Delta \over \Lambda_0} 
\right)^{2\alpha (0)/[1-2\alpha (0)]} 
\label{taweak} 
\end{eqnarray}  
which is a valid result even when $2 \alpha(0) \ln(\Lambda_0/\Delta) 
\sim {\cal O}(1)$ although its derivation requires $\alpha(0) \ll 1$. 
If, on the other hand, $\alpha(0) > h(0)$ then the  
$\alpha^2$ term dominates and the flow of $\alpha (l)$  
and we must take into account the $l$-dependence  
of $\alpha (l)$ in solving for  
the flow of $h(l)$  to strong coupling.  This leads to: 
\begin{eqnarray} 
T_A \approx \Delta (1+2 \alpha(0) \ln(\Lambda_0/\Delta))^{-1}. 
\label{tastrong} 
\end{eqnarray} 
Observe that (\ref{taweak}) and (\ref{tastrong}) are identical 
when $2 \alpha(0) \ln(\Lambda_0/\Delta) \ll 1$ but give a very 
different result when $2 \alpha(0) \ln(\Lambda_0/\Delta) \sim {\cal O}(1)$. 
We immediately notice that the $\alpha^2$ term in the 
RG destroys the KT transition. Unlike the Kondo problem the system  
retains coherence even at large coupling and is never overdamped. 
This is a quantum mechanical effect and 
comes from the fact that the spin operators do not commute. While 
the $S_1$ operator in (\ref{qfm}) wants to orient the impurity 
spin in its direction, the same happens for the $S_2$ operator. In a 
classical system (large $S$) the spin would orient in a finite angle 
in the XY plane. However for a finite $S$ impurity this is not possible 
and the impurity coupling is effectively {\it quantum frustrated} reducing 
the effective coupling to the environment. Another interesting feature of 
the RG flow is that for $h(\ell) \to 0$ we find, 
\begin{eqnarray} 
\alpha^*=\alpha(\ell^*) = \frac{\alpha(0)}{1+ 2 \alpha(0) \ell^*}  
\approx \frac{1}{2 \ln(\Lambda_0/T_A)} \, , 
\label{asta} 
\end{eqnarray}  
when $ 2 \alpha(0) \ell^* \gg 1$,  
$\alpha (l)$ is essentially independent of $\alpha(0)$ 
at energy scale $T_A$. While $T_A$ gives the crossover energy scale 
between weak and strong coupling, $\alpha^*$ provides information about the 
dissipation rate, $\tau^{-1}$, of the impurity dynamics. Our results indicate that  
for $\alpha(0) \ell^*$ sufficiently large, $\tau^{-1}$ is 
independent of the initial coupling to the bosonic baths. 
 
\begin{figure}[htbp] 
\includegraphics[%
  width=0.80\columnwidth]{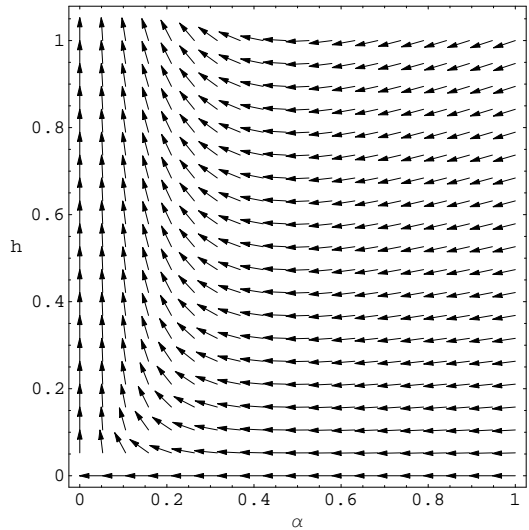} 
\caption{\label{isoflow}Renormalization group flow given by 
  eqs.~(\ref{rgiso}) in the $\alpha$ versus $h$ plane.} 
\end{figure}

In Fig.~\ref{isoflow} we depict the RG flow in the $\alpha$ versus 
$h$ plane. As discussed above, we can see that asymptotically 
(that is, large $\ell$) $\alpha(\ell)$ renormalizes to zero 
while $h(\ell)$ becomes large.   
An interesting feature of this RG, as we pointed out above, is that for large 
values of $\alpha(0)$ (large coupling to the environment) and intermediate 
values of $\ell$ the renormalization of $h(\ell)$  
becomes independent of $\alpha(0)$. This indicates that there is 
a single variable that determines the RG flow at intermediate energy 
scales. The fact that only one coupling determines the RG flow 
indicates that there must be {\it scaling} in the physical properties 
with the renormalized value of $h$. In the next section we 
will discuss how the RG results reflect on the behavior of the 
transverse susceptibility.

\section{Impurity Susceptibility \label{sec:susceptibility-at-$T=3D0$}} 
 
In the previous section we discusse the RG calculation in the  
weak coupling limit. The RG 
indicates that for large values of the couplings nothing new 
should happen. Nevertheless, given the perturbative nature of 
our analysis, this conclusion may not be warranted. Our  
conclusions can be put on firmer ground with the use of  
numerical renormalization group (NRG) \cite{nrg}. In NRG 
we do not look at the renormalization of the couplings, 
as we did in the previous section, but at the behavior  
of the susceptibility itself. Thus, in the first part 
of this section we study the behavior of the susceptibility 
as a function of the frequency at $T=0$ with NRG. In the second part  
of this section, based on the perturbative RG of the previous section 
and the NRG, we obtain analytic 
expressions for the transverse susceptibility in various regimes. 
We show that these two methods provide full 
support for the RG equations obtained in the previous section.

\subsection{Numerical Renormalization Group (NRG)} 
 
In order to learn more about the model we have performed numerical 
renormalization group (NRG) calculations \cite{nrg} 
on the Hamiltonian (\ref{qfm}).   
 Although NRG has recently been extended to bosonic models 
\cite{bulla03}, we follow a more traditional approach and  
transform (\ref{qfm}) into a fermionic problem.  
However, the bosonic baths $\phi_1$ and $\phi_2$ being Ohmic, we can 
also represent them as the spin density fluctuations of two fermion 
fields, $\psi_1$ and $\psi_2$, 
\begin{eqnarray} 
H_{F}&=& \frac{\Delta}{2} \sigma_3+\sum_{k,\mu,i=1,2} v_F k c^\dagger_{ik\mu} c_{ik\mu}\nonumber\\ 
&+& \frac{1}{2}g_{1} 
S_1\psi^\dagger_1\sigma_1\psi_1+\frac{1}{2}g_{2} 
S\psi^\dagger_2\sigma_2\psi_2\;,  
\label{eq:fermionicSB} 
\end{eqnarray} 
where $v_F$ is the Fermi velocity and  
$\psi_{i\mu}=\sum\limits_kc_{ik\mu}$ are the local 
fermion operators. Notice that we have two different 
set of fermions (labeled by $i=1,2$) that couple 
by x and y-component of their {}``spin'' to  
the corresponding  
components of the impurity spin.  
 
\begin{figure}[htbp] 
\includegraphics[%
  width=0.90\columnwidth]{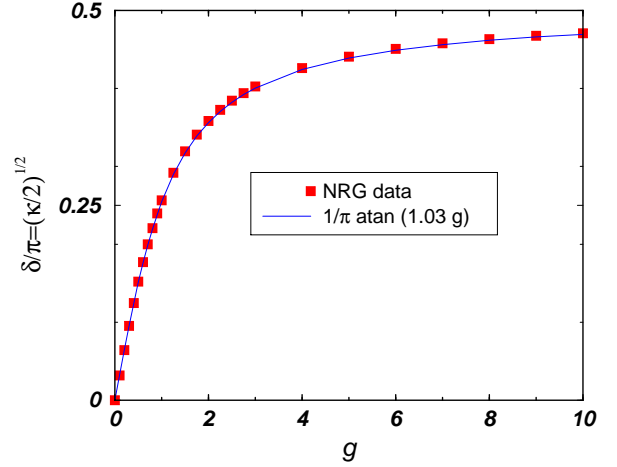} 
\caption{The phase shift  
(and therefore the bosonic coupling)  
extracted from the NRG finite size spectra as 
a function of the fermionic coupling.} 
\label{fig:calib} 
\end{figure}

In order for (\ref{eq:fermionicSB}) to be a faithful 
representation of (\ref{qfm}) one has to map the bosonic 
couplings $\alpha_{1,2}$ into the fermionic couplings 
$g_{1,2}$. As in the case of the Kondo problem \cite{ayh} 
the bosonic couplings are related to the electronic  
couplings through the electronic phase shifts  
$\delta_{1}$ and $\delta_{2}$: 
\begin{eqnarray} 
\alpha_i = 2\left(\frac{1}{\pi} \delta_i\right)^2\;. 
\end{eqnarray} 
Here the phase shifts can be determined directly from the NRG spectrum. 
The price what one has to pay for this simplicity 
is that the entire parameter space of the fermionic model $0\leq 
g_{i}\leq\infty$ covers only a smaller regime of the original 
model $0\leq\alpha_i\leq 1$ and therefore the localization 
transition is beyond the boundaries of the method.  
The phase shifts are given with a very good accuracy by: 
\begin{equation} 
\delta_i={\rm atan}(f(\Lambda_{\rm NRG})g_{i})\;, 
\end{equation} 
where $\Lambda_{\rm NRG}$ is the parameter of the logarithmic discretization 
used in NRG and $f(\Lambda_{\rm NRG})$ is a numerically determinable factor 
close to unity. For the numerical work we used 
$\Lambda_{\rm NRG}=2$  
and we find $f(\Lambda_{\rm NRG}=2)=1.03$ (see Fig.~\ref{fig:calib}). 

\begin{figure}[htbp] 
\includegraphics[%
  width=0.90\columnwidth]{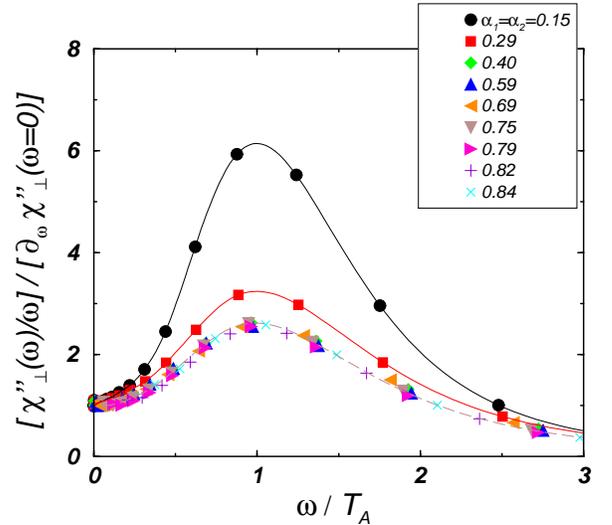} 
\caption{$\chi''_\perp(\omega)/\omega$ as a function of $\omega/T_A$.} 
\label{fig:fig1} 
\end{figure}

In Fig.\ref{fig:fig1} we show the results for  
$\chi^{''}_{\perp}(\omega)/\omega$  
(normalized to its value at $\omega=0$)  
as a function of $\omega/T_A$ (where $T_A$ is the crossover 
energy - see previous section) in the case 
when $\alpha_1=\alpha_2 =\alpha$ ($g_1=g_2$) as one varies $\alpha$. 
Notice that, in agreement with the RG calculation, the susceptibility 
retains a peak even for strong coupling indicating that the spin 
remains coherent. Furthermore, as the coupling increases the  
susceptibility curves collapse into a universal curve showing that 
at large couplings to the environment the susceptibility can be 
written in a scaling form: 
\begin{eqnarray} 
\chi^{''}_{\perp}(\omega,\alpha,h) = \chi_0 \, \, \, 
{\rm  f}\left(\frac{\omega}{T_A(\alpha,h)}\right) 
\end{eqnarray} 
where $\chi_0 = \partial_\omega\chi^{''}_{\perp}(\omega=0,\alpha,h)$ 
and ${\rm f}(x)$ is a universal function so that ${\rm f}(x \to 0) = x$ 
and ${\rm f}(x \to \infty) \approx 1/x$.  
These results are in agreement with our earlier conclusions based on the RG 
calculation. 

\begin{figure}[htbp] 
\includegraphics[%
  width=0.90\columnwidth]{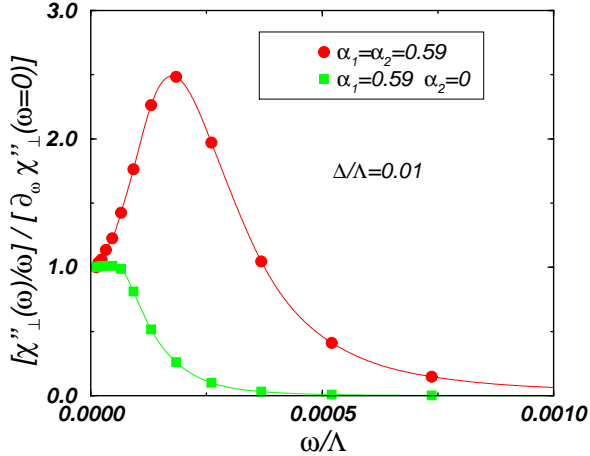} 
\caption{$\chi''_\perp(\omega)/\omega$ as a function of $\omega/\Lambda$.} 
\label{fig:fig2} 
\end{figure}

To compare results for our model with that of the single bath DTLS, 
we have calculated $\chi^{''}_{\perp}(\omega)$ for $\alpha_1=0.59$ and 
$\alpha_2 =0$ and compared with the case where $\alpha_1=\alpha_2=0.59$.  
The result is shown in Fig.\ref{fig:fig2}. Notice that in the DTLS case 
there is no trace of the peak in the susceptibility indicating that 
the relaxation of the spin is completely overdamped. However, in the 
isotropic case one finds a well defined peak even when the coupling 
to the environment is large, indicating that the spin still keeps 
memory of the tunneling splitting, even when strongly interacting with 
the bath. This is a clear demonstration of the effect of frustration 
of decoherence.

\subsection{Analytic Results} 
 
The RG results of section \ref{sec:perturbative-renormalization-group} 
show that the transverse couplings of the impurity to the 
environment always flow to $\alpha \to 0$ indicating that a  
perturbative approach should give a sensible result. When $\alpha=0$ the ground 
state of the problem is an eigenstate of $S_3$ and therefore the 
transverse susceptibility has a Dirac delta peak at $\omega=\Delta$, 
that is, zero relaxation rate, $\tau^{-1}=0$. In order to 
obtain a finite relaxation one makes use of the Bloch equations \cite{Abragam-01} 
for the expectation values of the spin operators, $M_i = \langle S_i \rangle$: 
\begin{eqnarray} 
\frac{d \vec{M}}{dt}= \frac{\Delta}{2} \vec{M} \times \vec{z} -  
\frac{M_1 \vec{x} +M_2 \vec{y}}{T_2} -\frac{M_3 \vec{z}}{T_1}, 
\nonumber 
\end{eqnarray} 
where $1/T_2$ is the transverse and $1/T_1$ is the longitudinal 
relaxation rates. It is straightforward to 
write a second order differential equation for $M_1(t)$: 
\begin{equation*} 
\frac{d^2 M_1}{dt^2} + \frac{2}{T_2} \frac{d M_1}{dt} +  
\left(\frac{\Delta^2}{4} +\frac{1}{T_{2}^2}\right) M_1 = 0 \, , 
\end{equation*} 
implying that the transverse correlation function has the form: 
\begin{eqnarray} 
\frac{\chi_{\perp}^{\prime\prime}\left(\omega\right)}{\omega} & 
\propto & 
\frac{2/T_2}{\left(\omega^{2}-\Delta^{2}/4 
    -1/T_2^{2}\right)^{2}+4\omega^2/T_2^{2}} \, . 
\label{eq:Bloch-sol} 
\end{eqnarray} 
In appendix~\ref{sec:RPA} we derive these results using a 
random phase approximation (RPA) and improve on them by  
replacing the bare 
values of the parameters by their renormalized RG value: 
\begin{eqnarray} 
\frac{\chi^{''}_{\perp}(\omega)}{\omega} =  
\frac{[\arctan(T_A \tau)]^{-1} T_A/\tau}{ 
\left(\omega^2-(T_A)^2-1/\tau^{2}\right)^2+4  
\omega^2/\tau^2} \, , 
\label{rpa} 
\end{eqnarray} 
where  
\begin{eqnarray} 
\tau^{-1} \approx \frac{\pi}{2} (\alpha^*)^2 T_A \, . 
\end{eqnarray}  
Notice that (\ref{rpa})  reduces to a Dirac delta function at  
$\omega=\Delta$ as $\alpha(0) \to 0$,  
as expected. We find that this approximation is good for  
$\omega \ll T_A$ and also describes well the NRG results for all  
$\omega <\Lambda_0$ when $\Lambda_0>T_A \gg \Lambda_0 \alpha^*$. 
In the zero frequency limit (\ref{rpa}) reduces to  
\begin{eqnarray} 
\chi^{''}_{\perp}(\omega=0) \approx (\alpha^*)^2 \omega/(T_A)^2 
+ {\cal O}[(\alpha^*)^4] 
\end{eqnarray} 
and the Kramers-Kronig relation immediately leads to real part 
of the susceptibility:  
\begin{eqnarray} 
\chi^{'}_{\perp}(\omega \approx 0)  
&=& \pi/[8 T_A (1+(\alpha^*)^4) \arctan(1/(\alpha^*)^2)] 
\nonumber 
\\ 
&\approx& 1/(4 T_A) + {\cal O}[(\alpha^*)^2]. 
\end{eqnarray}  

Although the RPA result gives good results in certain regimes it 
fails in the asymptotic cases.   
In those regimes a new approach has to be developed. For that purpose 
we will use the criteria of renormalizability of the theory in order 
to calculate the susceptibility. If we knew the exact $\beta$-functions 
of the theory,  
\begin{eqnarray} 
\beta_i(\{\alpha\}) = \frac{d \alpha_i}{d \ell}, 
\end{eqnarray}  
one could in principle integrate the exact RG flow in order to obtain 
the exact result. However, we only have access to the perturbative 
result (\ref{finalrg}) that indicates that there is no other 
fixed points in the problem. The question is whether these results 
are valid in other regimes.  
 
Let us consider some limiting cases of the problem at hand. 
Firstly consider the special situation where  
$\alpha \left(0\right)=\alpha _{1}\left(0\right)=\alpha _{2}\left(0\right)$ 
and there is no magnetic field, $\Delta=0$  
($h\left(0\right)=0$). 
In this case the Hamiltonian of the problem can be written, from 
(\ref{eq:hamil-0}), as: 
\begin{eqnarray} 
H_{\mathtt{eff}} & = & \int dx\sum _{\alpha =1,2}\left(\partial _{x}\phi 
  _{\alpha }(x)\right)^{2}-\sqrt{8\pi \alpha }\partial _{x}\phi _{1}(0)S_{1} 
\nonumber 
\\ 
 & - & \sqrt{8\pi \alpha }\partial _{x}\phi _{2}(0)S_{2}.  
\label{zerofield} 
\end{eqnarray} 
From the renormalization group equations, Eqs.~(\ref{rgiso}), we find:  
\begin{equation} 
\beta(\alpha) = \frac{d \alpha}{d\ell} = - 2 \alpha^{2}+{\cal O}[\alpha^{3}]. 
\label{RG-alpha} 
\end{equation} 
At finite temperature $T \ll \Lambda$ the RG flow is cut-off 
by the temperature and we can write $d\ell \approx - dT/T$ and 
use the temperature as the cut-off. We can solve (\ref{RG-alpha}) 
for $\alpha$ as a function of $T$ at once: 
\begin{equation} 
\alpha(T)\approx {\frac{\alpha_{0}}{1+2\alpha_{0}\ln (\Lambda_{0}/T)}} 
\label{aleff} 
\end{equation} 
When $T \to 0$ one finds:  
\begin{equation} 
\alpha(T) \approx {\frac{1}{2\ln (\Lambda_{0}/T)}}, 
\label{alas} 
\end{equation} 
which is independent of $\alpha(0)$ in agreement with (\ref{asta}).  
 
We first consider the susceptibility at finite $T$ and zero frequency, 
\begin{equation} 
\chi (T,\Lambda,\alpha_{0})={\frac{1}{4 T}} \, g(T/\Lambda,\alpha_{0}), 
\label{chig} 
\end{equation} 
where $g(x)$ is a dimensionless function. Since the theory is renormalizable,  
the susceptibility should obey the Callan-Symanzik (CS) equation \cite{amit}:  
\begin{equation} 
\left[\Lambda{\frac{\partial }{\partial \Lambda}}+\beta(\alpha){\frac{\partial 
    }{\partial \alpha}}+2\gamma(\alpha)\right]g(\alpha,T/\Lambda)=0 \, , 
\label{casy} 
\end{equation} 
where $\gamma(\alpha)$ is the anomalous dimension associated with the 
operator $\sigma_1$. Equation (\ref{casy}) expresses the fact that a change in  
the cut-off, $\Lambda$, can be exactly 
compensated for by a change in the bare coupling, $\alpha$, 
together with a rescaling of the susceptibility. 
The most general solution of (\ref{casy}) is:  
\begin{equation} 
g(\alpha,T/\Lambda)= \exp\left\{\int _{\alpha_{0}}^{\alpha (T)}[ 
2\gamma(\alpha)/\beta (\alpha)]d\alpha \right] \, h[\alpha(T)], 
\label{sol1} 
\end{equation} 
 where $h(x)$ is an arbitrary function of the renormalized coupling.  
We can rewrite (\ref{sol1}) in a slightly different form: 
 \begin{equation} 
g(T)=\exp \left[\int _{\alpha_{0}}^{1}[2\gamma (\alpha )/\beta (\alpha 
  )]d\alpha \right]\, \, \phi [\alpha (T)], 
\label{sol2} 
\end{equation} 
where we have introduce a new function $\phi[\alpha]$ and used that  
\begin{equation} 
\exp \left[\int_{1}^{\alpha(T)}[2\gamma (\alpha )/\beta (\alpha )]d\alpha 
\right] 
\nonumber 
\end{equation} 
is by itself some (in general unknown) function of $\alpha (T)$ and 
we have absorbed this term into the function $h(\alpha (T))$.  
Hence a non-zero anomalous dimension implies some residual explicit dependence 
of $g(T)$, and hence $\chi (T)$, on the bare coupling $\alpha_{0}$ 
in addition to its implicit dependence on $\alpha_0$ through 
the renormalized coupling.  
Notice from (\ref{alas}) that $\alpha (T)$ becomes small 
at low $T$ and therefore one can expand $\phi (\alpha (T))$ 
in a power series in $\alpha (T)$. In this case, replacing (\ref{sol2}) 
into (\ref{chig}) we find: 
\begin{equation} 
\chi (T)={\frac{1}{4T}}\exp \left[\int _{\alpha_{0}}^{1}[2\gamma (\alpha )/ 
\beta (\alpha)]d\alpha \right] \sum _{n}\, b_{n} \, \alpha^{n}(T), 
\label{finalchi} 
\end{equation} 
where $b_n$ are the coefficients of the expansion of $\phi[\alpha]$. 
 
Eq.~(\ref{finalchi}) is formally exact. However, one does not know 
the anomalous dimension a priori. One way to go about this is to 
compare the exact result (\ref{finalchi}) with the perturbative 
result obtained in leading order in $\alpha_0$. In appendix  
\ref{sec:Perturbation-Theory} we show that perturbation theory gives: 
\begin{equation} 
\chi (T)={\frac{1}{4T}}[1+2\alpha_{0}\ln (T/\Lambda_0)+ 
{\cal  O}[\alpha_{0}^{2}] \, . 
\label{chip} 
\end{equation} 
Replacing (\ref{alas}) in (\ref{finalchi}) and  
comparing with (\ref{chip}) we find that $b_{0}=0$, $b_{1}=1$ 
and  
\begin{equation} 
\int _{\alpha_{0}}^{1}[2\gamma (\alpha )/\beta (\alpha ) d\alpha] \approx  
-\ln(\alpha_{0}) \, . 
\end{equation} 
The coefficient of $2$ in front of the second term in (\ref{chip})  
is crucial. Note that what fixes the definition of $\alpha$ is the RG 
equation, Eq. (\ref{RG-alpha}). Since $\beta(\alpha) \approx - 2 \alpha^{2}$, 
we see that:  
\begin{equation} 
\gamma (\alpha)= - \alpha +{\cal O}[\alpha^{3}] \, , 
\end{equation} 
is the value of the anomalous dimension in leading order in $\alpha$. 
Therefore, we have concluded that 
\begin{equation} 
\chi (T) \approx {\frac{1}{8T\alpha_{0} \ln(\Lambda_0/T)}} \, , 
\end{equation} 
when $T \to 0$. This result is expected to be true even at very low $T$  
when $\alpha_{0}\ln (\Lambda/T) \geq 1$. 
Suppose that the bare coupling, $\alpha_{0}$ is not small. What 
can we say from the RG in this case? As long as we consider very low 
$T$ where $\alpha (T)\ll 1$ so that $\phi (\alpha (T))\approx \alpha (T)$, 
we have:  
\begin{equation} 
\chi (T)\approx \exp \left[\int _{\alpha_{0}}^{1}[2\gamma (\alpha )/ 
\beta (\alpha)\right]{\frac{1}{8T\ln (\Lambda_0/T)}}. 
\label{chibt} 
\end{equation} 
The first factor is some unknown function of the bare coupling 
but the $T$ dependence is the same as before.  
 
Now consider the susceptibility at $T=0$ but finite frequency.  
Once again, thanks to the renormalizability of the theory  
$\chi ''(\omega )$ obeys the 
same CS equation with the same $\beta $-function and the same anomalous 
dimension, $\gamma (\alpha)$. This anomalous dimension is  
a property of the spin operator $\sigma_1$ and  
must be the same for either finite $T$ and $\omega=0$ or finite $\omega$ 
and $T=0$. Therefore, following the earlier discussion it must have the form:  
\begin{equation} 
\chi ''(\omega ,\Lambda,\alpha_{0})= 
\exp \left[\int _{\alpha_{0}}^{1}[2\gamma (\alpha)/\beta (\alpha)\right]{ 
\frac{1}{\omega}}\, F(\alpha(\omega )). 
\label{chiw} 
\end{equation} 
The function $F(\alpha(\omega ))$ is not necessarily the same as  
$\phi (\alpha (T))$ and, in general, is unknown.  
However, the first factor, giving the 
explicit dependence on $\alpha_{0}$ in a perturbative expansion,  
should be exactly the same as in the previous calculation of $\alpha(T)$.  
Thus, if $\alpha_{0}\ll 1$, we must have:  
\begin{equation} 
\chi ''(\omega ,\Lambda,\alpha _{0})={\frac{1}{\omega \alpha_{0}}} \,  
F(\alpha(\omega )). 
\label{chisa} 
\end{equation} 
Again, we perform ordinary perturbation theory for $\chi^{''}(\omega)$  
in powers of $\alpha_{0}$ and improve the perturbative result with the RG 
by matching it to (\ref{chisa}) by expanding in powers of $\alpha(\omega)$.  
Since we already know $\beta (\alpha)$ and $\gamma (\alpha)$, 
the result must have a rather restricted form to be a solution of 
the CS equation. The susceptibility at finite frequency is given by 
Eq.~(\ref{eq:xi-T0}): 
\begin{equation} 
\chi^{\prime \prime }(\omega ,\Lambda,\alpha_{0}) \approx { 
\frac{1}{\omega }}\left[\alpha_{0}-4\alpha_{0}^{2} \ln (\Lambda_0/\omega ) 
+ {\cal O}(\alpha_0^3) \right]. 
\end{equation} 
This result is consistent with the RG form of Eq. (\ref{chisa}) if we assume 
that:  
\begin{eqnarray} 
F(\alpha (\omega )) & \approx  & \alpha^2(\omega) \, , 
\label{fkappa} 
\\ 
 & \approx  & \alpha_{0}^{2}-4\alpha_{0}^{3}\ln (\Lambda_0/\omega ). 
\nonumber 
\end{eqnarray} 
Having found the function $F(\alpha(\omega ))$ at small $\alpha(\omega )$ 
we can now invoke the RG. In particular, for small bare 
coupling and small $\omega $ we have:  
\begin{equation} 
\chi^{\prime \prime}(\omega) \approx {\frac{1}{4\alpha_{0}\omega  
\ln^{2}(\Lambda_0/\omega)}}. 
\label{chippw} 
\end{equation} 
Even if the bare coupling is not small, but we go to small enough 
$\omega$ so that $\alpha (\omega )\ll 1$, the RG implies that:  
\begin{equation} 
\chi^{\prime \prime }(\omega) \approx  
\exp \left[\int_{\alpha_{0}}^{1}2\gamma (\alpha )/\beta (\alpha) d\alpha\right] 
{\frac{1}{4\omega \ln ^{2}(\Lambda_0/\omega )}}, 
\end{equation} 
where the first term in an  unknown function of the bare coupling constant. 
Thus, eqs.~(\ref{chibt}) and (\ref{chippw}) give the 
temperature and frequency behavior of the susceptibility for, $\alpha^* \ll 
1$, $T_A \ll \Lambda_0 \alpha^*$, and   
in the frequency and temperature range $T_A \ll \omega, T \ll \Lambda_0$. 
When these conditions are satisfied the ratio, 
\begin{eqnarray} 
\frac{\chi^{\prime \prime}(\omega,T=0)}{\chi(T)} 
= \frac{2 T}{\omega} \frac{\ln(\Lambda_0/T)}{\ln^2(\Lambda_0/\omega)} \, , 
\label{uniratio} 
\end{eqnarray} 
is universal.  
 
\section{Discussion and Conclusions \label{conclusions}} 
 
Decoherence can be defined as the unavoidable evolution 
of the total state of the system and the environment towards an  
entangled state. 
This is a dynamical definition of decoherence, and clearly shows 
the conceptual difference between dissipation (that 
involves the transfer of energy from the subsystem to the environment) 
and decoherence. 
An important concept in the study of decoherence is the notion of 
a \emph{preferred basis}: every time that a system interacts with 
an environment, a set of states is naturally selected by the form of 
the interaction.  
The text book example is given by the exactly solvable model\cite{W99,BHP02} 
\begin{eqnarray*} 
H_{deco} & = & \sum_{k>0} k a_{k}^{\dagger}a_{k} 
  +  i\frac{\lambda_{1}}{\sqrt{2 L}} S_{1}\sum_{k>0}\sqrt{k}(a_{k}-a_{k}^{\dagger}). 
\end{eqnarray*} 
Although this model does not have any dissipative mechanism, the two level 
system experience strong decoherence. Suppose that the system is prepared 
at time $t=0$ as a direct product of the bath and the two level system: 
\begin{eqnarray*} 
\rho (t=0) & = & \rho_{bath}(t=0)\otimes \rho_{\vec{S}}(t=0), 
\end{eqnarray*} 
where $\rho_{bath}$ and $\rho_{\vec{S}}$ are respectively the density 
matrices  of bath and the two level system.  
A natural basis choice for the two level system is $S_{1}$. If we further 
suppose  
that the two level system is prepared in a state of $S_{2}$, then at $t=0$  
the reduced density matrix,  
\begin{eqnarray*} 
\rho_R (t=0) & = & tr_{bath}\left[\rho_{bath}(t=0)\otimes \rho_{\vec{S}}(t=0)\right], 
\end{eqnarray*} 
has off-diagonal matrix elements indicating that the system is coherent.  
As the system evolves in time the off-diagonal elements decay very fast  
due to the entanglement of $\vec{S}$ and the bath degrees of freedom.  
As $t\to\infty$ only the diagonal elements of the reduced density matrix  
of the system remain, and we say that the system {}``decoheres'' 
to the \emph{preferred basis} of $S_{1}$. 
 
With this example in mind is now simple to understand the effects that  
we described in this manuscript. Consider the Hamiltonian (\ref{qfm}). 
In this case it is no longer possible to define a \emph{preferred basis} for the 
two level system. The entanglement of $\vec{S}$ with each one of the 
baths is suppressed by the other, and as a result the decoherence phenomena 
is frustrated. This physical picture shows the true meaning of our results,  
the {}``quantum frustration'' is the lack of a \emph{preferred basis} for the  
system of interest.  
 
The quantum frustration of decoherence can be also understood as a  
result of a version of Coleman's or Mermin-Wagner's theorem \cite{coleman}.   
When $\alpha_1=\alpha_2$ there is a $U(1)$ symmetry in impurity problem.   
Hence, one has an effective $(1+1)$ dimensional field theory with $U(1)$ 
symmetry so this symmetry cannot be spontaneously broken even at $T=0$. 
In fact, because one has a single boundary degree of freedom, one can also think 
of the problem as an almost $(0+1)$ dimensional field theory. So it is a rather 
remarkable fact that even the $Z_2$ symmetry which remains when $\alpha_2=0$ 
can be spontaneously broken, as in the case of the DTLS. The $U(1)$ symmetry would  
have to be spontaneously broken 
 in a phase in which the spin is localized in an eigenstate of either  
$S_1$ or $S_2$. Quantum mechanics prevents that from happening.  
 
One can ask how generic this result really is. 
For quantum frustration to occur the coupling  
constants with all the baths must be identical. This can be achieved  
when the role of the two baths is played by two Goldstone modes, 
resulting from the spontaneous breaking of a continuous symmetry, 
such that the residual unbroken symmetry rotates the two Goldstone 
modes into each other. 
When the couplings are not exactly equal quantum frustration occurs up 
to a certain energy scale below which one of the heat baths takes over 
and one obtains the standard decoherence problem in dissipative ohmic 
systems. In terms of Fig.\ref{cap:RGflow-1} it means that the asymptotic 
flow is the one for either $\alpha_1=0$ or $\alpha_2=0$.  
In summary, quantum frustration of decoherence is a general phenomena.  
It has clear implications to quantum/classical transition and measure 
theory.  
Moreover it is potentially important to the development of technologies where  
decoherence is a fundamental issue as in the case of quantum communication  
and quantum computation. 
 
In summary, we have studied a model of quantum frustration of decoherence 
in open systems. Contrary to standard dissipative models with ohmic 
dissipation, the non-commutative nature of spin operators lead to a 
frustration of decoherence. We have shown that while in a DTLS the 
spin dynamics becomes overdamped at large couplings with a heat bath, 
in a system with quantum frustration it is always underdamped and the 
system keeps the memory of its quantum nature. Using perturbative RG 
calculations we have shown that at large couplings with the bath the 
transverse spin susceptibility shows scaling with a characteristic energy 
scale $T_A$, the analogous of the Kondo temperature in the DTLS, that 
separates the region of strong to weak coupling. We have supported our 
claims with NRG calculations and have calculated the frequency and 
temperature dependence of the transverse susceptibility using the 
renormalizability of the theory. Our results may be applicable to 
a large class of problems where decoherence plays a fundamental role.

\begin{acknowledgments} 
The authors would like to thank C.~Chamon, and F.~Guinea.  A.H.C.N. was 
partially supported through NSF grant DMR-0343790. L.~B. and G.~Z.  
acknowledge support by Hungarian Grants No. OTKA T038162, T046267, D048665, 
and T046303, and the European 'Spintronics' RTK HPRN-CT-2002-00302. 
\end{acknowledgments}

\appendix 
 
\section{Impurity spin in a magnetically ordered environment\label{sec:Impurity-spin-in}} 
 
In this appendix we will show how quantum frustration can arise in the 
context of a magnetic impurity in a magnetic environment \cite{prl}.  
Let us consider a magnetic environment describe by the quantum Heisenberg 
Hamiltonian in the presence of an impurity: 
\begin{eqnarray} 
H & = & J\sum_{\langle i,j \rangle} 
\vec{s}_{i} \cdot \vec{s}_{j} + \lambda \, \vec{S} \cdot \vec{s}_{0}, 
\label{eq:model-1} 
\end{eqnarray} 
where $J$ is the magnetic exchange between nearest neighbor  
spins $\vec{s}_i$ located 
on a lattice site $\vec{R}_i$ in $d$ dimensions, $\lambda$ is the 
coupling between the environmental spins and an impurity spin  
$\vec{S}$ located at the origin of the coordinate system.  
In what follows we will consider the antiferromagnetic case of 
$J>0$ although the ferromagnetic case ($J<0$) can be studied 
in an analogous way.  
 
The partition function of the problem in spin coherent state path integral 
can be written as \cite{fradkin}: 
\begin{eqnarray*} 
Z & = & \int D\vec{N}\,\delta(\vec{N}^{2}-1) \int D\vec{n} \, 
\delta(\vec{n}^{2} -1)\,\, 
e^{-iS_{\mathtt{B}}(\vec{N})-S(\vec{N},\vec{n})} 
\end{eqnarray*} 
where $\vec{N}$ represents the impurity spin and $\vec{n}(\vec{r})$ the 
environmental spin field,  $S_{\mathtt{B}}$ is the Berry's phase,  
\begin{eqnarray} 
S &=& \int d\tau d\vec{r} \left\{ \frac{1}{2g} 
\left[ 
\left(\partial_{\tau}\vec{n}(\tau,\vec{r})\right)^{2} 
+c^{2}\left( \nabla \vec{n}(\tau,\vec{r})\right)^{2} 
\right]  
\right. 
\nonumber 
\\ 
&+& \left. \lambda \delta\left(\vec{r}\right) \vec{n}(\tau,\vec{r})  
\cdot \vec{N}\left(\tau\right)\right\} , 
\label{fullaction} 
\end{eqnarray} 
is the action of the problem where  
$g=c^{2}/\rho_{s}$ is the coupling constant ($c=2\sqrt{d}Jas$ 
is the spin wave-velocity and $\rho_{s}=Js^{2}a^{2-d}$ is the spin 
stiffness, $a$ is the lattice spacing and $s$ is the value of the 
environmental spin).  
 
Assume that the $O(3)$ symmetry of the model is spontaneously broken 
so that the field $\vec{n}$ orders. In this case we can write: 
\begin{eqnarray} 
\vec{n}\left(\tau,\vec{r}\right) & \approx & \left(\varphi_{1}\left(\tau,\vec{r}\right),\varphi_{2}\left(\tau,\vec{r}\right),1\right), 
\label{eq:vecfield} 
\end{eqnarray} 
where $\varphi_{1,2}$ are small fluctuating fields corresponding 
to the two Goldstone modes of the antiferromagnet.  
A possibility that is not considered in this work is associated 
with the formation of a spin texture around the impurity spin. 
In a classical spin system a spin texture can be formed in the 
bulk spins due to the presence of strong and/or anisotropic interactions.  
The spin texture can follow the impurity as it tunnels 
invalidating the methods used here (an instanton 
calculation is required to take into account the collective nature 
of the texture). The results in this appendix are only valid if 
no spin texture is formed around the magnetic impurity. 
 
In the ordered 
phase the Berry's phase term is unimportant and can be dropped.  
Using Eq.~(\ref{eq:vecfield}) the action (\ref{fullaction}) reads, 
\begin{eqnarray} 
S & \approx & \sum_{\alpha=1,2}\int d\tau d\vec{r}\left\{ \frac{1}{2g}\left[ 
\left(\partial_{\tau}\varphi_{\alpha}(\tau,\vec{r})\right)^{2} 
+c^{2}\left(\nabla\varphi_{\alpha}(\tau,\vec{r}\right)^{2}\right] 
\right. 
\nonumber 
\\ 
&+& \left. \lambda\delta\left(\vec{r}\right)\varphi_{\alpha}(\tau,\vec{r})\cdot 
N_{\alpha}(\tau) 
+\lambda N_{3}\left(\tau\right)\right\}. 
\label{eq:action-2} 
\end{eqnarray} 
We see that the action for the fields $\varphi_{1,2}$ is quadratic and 
therefore these fields can be traced out of the problem exactly. In this 
case the effective action for the impurity spin becomes in Fourier space: 
\begin{eqnarray*} 
S_{\mathtt{eff}} & \approx & \frac{g\lambda^{2}}{2}\sum_{\alpha=1,2} 
\int d\omega d\vec{k}\frac{N_{\alpha}(\omega)N_{\alpha}(-\omega)}{\omega^{2}+c^{2} {k}^{2}}\\ 
 & + & \lambda\int d\tau N_{3}\left(\tau\right). 
\end{eqnarray*} 
As we should expect from the spherical symmetry of the problem, the 
angular dependence in $\vec{k}$ can be integrated and we finally 
obtain 
\begin{eqnarray} 
S_{\mathtt{eff}} &\approx&  \frac{\alpha^{2}}{\Gamma(d-1)} \sum_{\alpha=1,2} 
\int_{-\infty}^{+\infty} d\omega \left[\int_0^{\infty} dq\, 
\frac{q^{d-1}}{\omega^{2}+q^{2}}\right] N_{\alpha}(\omega)N_{\alpha}(-\omega)\, 
\nonumber 
\\ 
 & + & \lambda\int d\tau N_{3}\left(\tau\right) \, , 
\label{sint} 
\end{eqnarray} 
where $q=c k$, and  
\begin{eqnarray*} 
\alpha^{2} = \frac{g \Gamma(d-1) \lambda^{2}}{2^{d+2}\pi^{\left(d-2\right)/2}\Gamma\left(d/2\right)c^{d}}, 
\end{eqnarray*} 
where $\Gamma(x)$ is a gamma function. 
Integrating (\ref{sint}) over $q$ and Fourier transforming back the 
frequencies to imaginary time we find: 
\begin{eqnarray} 
S_{\mathtt{eff}} & \approx & 
\alpha^{2} \sum_{\alpha=1,2} 
\int d\tau \int d\tau' 
\frac{N_{\alpha}(\tau)N_{\alpha}(\tau')}{|\tau-\tau'|^{d-1}} 
\nonumber  
\\ 
 & + & \lambda\int d\tau N_{3}\left(\tau\right), 
\label{eq:tracedaction} 
\end{eqnarray} 
which shows that the impurity interact in imaginary time through a long-range 
interaction that decays like $1/\tau^{d-1}$. 
 
The action (\ref{eq:tracedaction}) can be simplified by introducing a 
Hubbard-Stratanovich field that splits the interaction term. This can 
be done with the introduction of one-dimensional bosonic fields defined as:  
\begin{eqnarray*} 
\phi_{\alpha}\left(x,\tau\right) & = & 
\frac{T}{\sqrt{2L}}\sum_{q>0}\sum_{\omega_{n}} 
\frac{e^{iqx+i\omega_{n}\tau}}{\sqrt{q}}\phi_{\alpha}^{*}\left(q,\omega_{n}\right) 
\\ 
 & + & 
 \frac{e^{-iqx+i\omega_{n}\tau}}{\sqrt{q}}\phi_{\alpha}\left(q,\omega_{n}\right), 
\end{eqnarray*} 
$L\to\infty$ is the size of the one-dimensional line.  
Using these new fields the action (\ref{eq:tracedaction}) can be written as: 
\begin{eqnarray} 
S_{\mathtt{eff}} & = & 
\sum_{\alpha=1,2} T \sum_{q>0}\sum_{\omega_{n}} 
\left[i\omega_{n}+q\right]\phi_{\alpha}^{*}\left(q,\omega_{n}\right)\phi_{\alpha}\left(q,-\omega_{n}\right) 
\nonumber  
\\ 
 & + & \alpha\left|q\right|^{\frac{d-1}{2}}\left[\phi_{\alpha}^{*} 
\left(q,\omega_{n}\right)N_{\alpha}(\omega_{n})+\phi_{\alpha}\left(q,\omega_{n}\right)N_{\alpha}(-\omega_{n}) 
\right] 
\nonumber  
\\ 
 & + & \lambda\int d\tau N_{3}\left(\tau\right). 
\label{eq:bosonicactiond} 
\end{eqnarray} 
It is easy to see that the trace over the bosonic fields reproduces (\ref{eq:tracedaction}). 
It is straightforward to see that in $d=3$ the above action reduces to (\ref{eq:hamil-0}).

\section{RG equations} 
\label{rgeq} 
 
In this appendix we will derive the RG equations (\ref{eq:rg-1}). From 
equation (\ref{vertices1}) we have: 
\begin{eqnarray} 
\langle A_{1}^{\pm} \rangle_> & = & \langle e^{\mp 
  i\sqrt{8\pi\alpha_{1}}(\phi_{1,<}\left(x=0\right)+\phi_{1,>}\left(x=0\right)) }S^{\pm} \rangle_> 
\nonumber 
\\ 
&=&  A_{1,<}^{\pm} \langle e^{\mp 
  i\sqrt{8\pi\alpha_{1}}\phi_{1,>}\left(x=0\right)} \rangle_> 
\nonumber 
\\ 
&=&  A_{1,<}^{\pm} e^{-\alpha_1 d\ell} \approx   
A_{1,<}^{\pm} (1-\alpha_1 d\ell)\, . 
\label{slow_fast_1} 
\end{eqnarray}  
Substituting (\ref{slow_fast_1}) in (\ref{eq:actionkappa1}) one obtains 
a term of the form: 
\begin{eqnarray} 
\delta \tau e^{d\ell} \Delta e^{-\alpha_1 d\ell} A_{1,<}^{m_j} =  
\delta \tau \Delta(\ell + d\ell) A_{1,<}^{m_j} 
\, , 
\end{eqnarray} 
where we have used that, by rescaling $\omega \to \omega/b$ (with  
$b = e^{d \ell} \approx 1 + d\ell$) one has  $\tau \to b \tau$. 
Hence, 
\begin{eqnarray} 
\Delta(\ell + d\ell) = \Delta(\ell) [1 + (1-\alpha_1) d\ell] \, , 
\end{eqnarray} 
and defining the dimensionless coupling, $h(\ell) = \Delta(\ell)/\Lambda$, 
one obtains (\ref{eq:rg-1-k3}). 
 
Analogously, from (\ref{vertices2}) we have: 
\begin{eqnarray} 
\langle B_{1}^{\pm} \rangle_> & = &  B_{1,<}^{\pm} e^{- d\ell} 
\langle e^{\mp  i\sqrt{8\pi\alpha_{1}}\phi_{1}\left(x=0\right)} \rangle_>  
\nonumber 
\\ 
&=& B_{1,<}^{\pm} e^{-(1+\alpha_1) d\ell}  
\nonumber 
\\ 
&\approx&  B_{1,<}^{\pm}  [1-(1+\alpha_1) d\ell] \, , 
\label{slow_fast_2} 
\end{eqnarray} 
where we have used that $\partial_x \to b^{-1} \partial_x$ since 
$k \to k/b$.  Replacing (\ref{slow_fast_2}) into the second 
term in the r.h.s. of (\ref{eq:actionkappa1}): 
\begin{eqnarray} 
\delta \tau \sqrt{\alpha_{2}} e^{-\alpha_1 d\ell} 
B_{1,<}^{m_{j}}\left(\tau_{j}\right) 
= \delta \tau \sqrt{\alpha_2(\ell+d\ell)} B_{1,<}^{m_{j}} \, , 
\end{eqnarray} 
and hence we write: 
\begin{eqnarray} 
\alpha_2(\ell + d\ell) = \alpha_2(\ell) e^{-2 \alpha_1 d\ell}  
\approx \alpha_2(\ell) (1-2 \alpha_1 d\ell) \, , 
\end{eqnarray} 
leading to equation (\ref{eq:rg-1-k2}).

\section{Perturbation Theory\label{sec:Perturbation-Theory}} 
 
In this appendix 
we show how to derive the perturbative expansion for the transverse 
susceptibility:  
\begin{eqnarray*} 
\mathcal{S}\left(\tau\right) & = & \left\langle 
  T_{\tau}  S_1\left(\tau\right)S_1\left(0\right)\right\rangle \, . 
\end{eqnarray*} 
 
\subsection{Static susceptibility $\omega=0$ and $h=0$} 
 
Firstly, let us consider the case of arbitrary $\alpha_2$ but small 
$\alpha_1$ (the case of arbitrary $\alpha_1$ and $\alpha_2 \ll 1$ is 
completely analogous). This regime can be obtained by using  
eq.~(\ref{rotation}):  
\begin{eqnarray*} 
\bar{H_{2}} & = & U_{2}^{-1}H_{\mathtt{eff}}U_{2} 
\\ 
 & = & 
 H_{0}-\sqrt{2\pi}\alpha_{1}\partial_{x}\phi_{1}e^{-i\sqrt{8\pi\alpha_{2}}\phi_{2}\left(0\right)}S^{+}+\mathtt{h}.\mathtt{c}. 
\end{eqnarray*} 
In this rotated basis, $\mathcal{S}\left(\tau\right)$ has a simple 
form: 
\begin{eqnarray*} 
\mathcal{S}\left(\tau\right) & = & \frac{1}{4} \left\langle 
    T_{\tau} \left(A_{2}^{+}\left(\tau\right)+A_{2}^{-}\left(\tau\right)\right) 
\left(A_{2}^{+}\left(0\right)+A_{2}^{-}\left(0\right)\right)\right\rangle , 
\end{eqnarray*} 
where\begin{eqnarray*} 
A_{2}^{\pm}\left(\tau\right) & = & e^{\mp 
  i\sqrt{8\pi\alpha_{2}}\phi_{2}\left(0,\tau\right)}S^{\pm}\left(\tau\right). 
\end{eqnarray*} 
The leading order terms in an expansion in powers of  
$\alpha_{1}$ at $T=0$ can be immediately 
obtained from the bosonic propagator: 
\begin{eqnarray} 
\mathcal{S}\left(\tau\right) & \approx &  
\frac{1}{4}\left|D\tau\right|^{-2\alpha_{2}} + {\cal O}[\alpha_1 \alpha_2] \, , 
\label{eq:correlationk2} 
\end{eqnarray} 
where $D$ is a short time cut-off.  
We can use the standard conformal transformation to promote this result to a  
finite temperature expression, 
\begin{eqnarray*} 
\mathcal{S}\left(\tau\right) \approx \frac{1}{4}\left(\frac{D}{T \pi} 
\sin\left|\pi T \tau \right|\right)^{-2\alpha_{2}} 
+ {\cal O}[\alpha_1 \alpha_2] \, . 
\end{eqnarray*} 
Expanding the above expression for $\alpha_{2}\ll 1$ gives: 
\begin{eqnarray*} 
\mathcal{S}\left(\tau\right) \approx \frac{1}{4}\left[1-2\alpha_{2} 
\ln\left(\frac{D}{T \pi}\sin\left|\pi\tau T\right|\right) 
+{\cal O}[\alpha_1 \alpha_2,\alpha_{2}^{2}] \right] . 
\end{eqnarray*} 
To this order, the susceptibility can be calculated immediately:  
\begin{eqnarray} 
\chi\left(T\right) & = & 
\int_{0}^{1/T}d\tau\mathcal{S}\left(\tau\right) 
\nonumber  
\\ 
 & \approx & \frac{1}{4 T}\left[1-\frac{2\alpha_{2}}{T} 
   \ln\left(\frac{D}{2\pi T}\right)+{\cal O}[\alpha_1 
   \alpha_2,\alpha_{2}^{2}] \right] \, . 
\label{eq:finitetemperaturexi} 
\end{eqnarray} 
  
For completeness, let us re-obtain this result by a direct perturbative  
calculation in second order in $\alpha_{2}$: 
\begin{widetext} 
\begin{eqnarray*} 
\mathcal{S}\left(\tau\right) & \approx & 
8\pi 
\alpha_2 \int_{\tau}^{1/T}d\tau_{1}\int_{0}^{\tau}d\tau_{2}\left[\left\langle 
    T_{\tau}\sigma_{2}\left(\tau_{1}\right)\sigma_{1}\left(\tau\right)\sigma_{2}\left(\tau_{2}\right)\sigma_{1}\left(0\right)\right\rangle  
- \left\langle 
     T_{\tau}\sigma_{1}\left(\tau\right)\sigma_{1}\left(0\right)\right\rangle 
   \left\langle 
     T_{\tau}\sigma_{2}\left(\tau_{1}\right)\sigma_{2}\left(\tau_{2}\right)\right\rangle \right]\left\langle T_{\tau}\partial\phi_{2}\left(\tau_{1}\right)\partial\phi_{2}\left(\tau_{2}\right)\right\rangle  
\\ 
 & = & 
 -\frac{\pi\alpha_2}{8}\int_{\tau}^{1/T}d\tau_{1}\int_{0}^{\tau}d\tau_{2}\left\langle \partial\phi_{2}\left(\tau_{1}\right)\partial\phi_{2}\left(\tau_{2}\right)\right\rangle . 
\end{eqnarray*} 
\end{widetext} 
Using the finite temperature propagator, 
\begin{eqnarray*} 
\left\langle 
  \partial\phi_{2}\left(\tau_{1}\right)\partial\phi_{2}\left(\tau_{2}\right)\right\rangle 
  & = & \frac{1}{4\pi}\frac{1}{\left[\frac{1}{\pi T}\sin\left[\pi T \left|\tau_{2}-\tau_{1}\right|\right]\right]^{2}}, 
\end{eqnarray*} 
we obtain 
\begin{eqnarray*} 
\mathcal{S}\left(\tau\right) & \approx & 
-\frac{\alpha_{2}}{4} \int_{\tau+\frac{1}{D}}^{1/T} d\tau_{1} \int_{0}^{\tau} 
d\tau_{2} 
\frac{1}{\left[ \frac{1}{\pi T} \sin\left[\pi T \left|\tau_{2}-\tau_{1}\right|\right]\right]^{2}} 
\\ 
 & \approx & 
 \frac{\alpha_{2}}{2}\left[\ln\left(\frac{\pi 
       T}{D}\right)-\ln\left(\sin\left|\pi T \tau\right|\right)\right] 
+{\cal O}[\alpha_1 \alpha_2, \alpha_2^2] \, , 
\end{eqnarray*} 
in agreement with (\ref{eq:finitetemperaturexi}).

\subsection{Dynamic susceptibility $\omega\neq0$ at $T=0$ and $h=0$} 
 
The finite frequency calculation is a little more tedious than the 
previous one. We would like to obtain the correlation function in 
fourth order in the coupling constants. From Eq.~(\ref{eq:correlationk2}) 
we already know part of the result, 
\begin{eqnarray} 
\mathcal{S}\left(\tau\right) =  
\frac{1}{4}-\frac{1}{2}\alpha_{2}\ln\left|D\tau\right|+\frac{1}{2}\alpha_{2}^{2} 
\ln^{2}\left|D\tau\right|+{\cal O}[\alpha_1 \alpha_2, \alpha_{2}^{3}]. 
\label{eq:cork2O4} 
\end{eqnarray} 
The remaining contribution to the correlation function is a term 
proportional to $\alpha_{1}\alpha_{2}$. A convenient way 
to derive this contribution is to use Eq.~(\ref{eq:rotation-1}) 
and compute the result to all orders in $\alpha_1$ but for $\alpha_2 \ll 1$. 
In second order in $\alpha_{2}$ we need to calculate: 
\begin{widetext} 
\begin{eqnarray*} 
\delta \mathcal{S}\left(\tau\right) & = & 
\frac{1}{4}+2\pi\alpha_{2}\int d\tau_{1}d\tau_{2}
\left\langle T_{\tau}S_3\left(\tau\right)B_{1}^{+} 
\left(\tau_{1}\right)B_{1}^{-}\left(\tau_{2}\right)S_3\left(0\right)\right\rangle 
-\pi\alpha_{2}\int d\tau_{1}d\tau_{2}\left\langle T_{\tau}B_{1}^{+} 
\left(\tau_{1}\right)B_{1}^{-}\left(\tau_{2}\right)\right\rangle . 
\end{eqnarray*} 
\end{widetext} 
From this point, it is straightforward to obtain the correlation function 
and the susceptibility at finite frequency: 
\begin{eqnarray*} 
\delta \mathcal{S}\left(\tau\right) & = & 
\frac{1}{4}-\frac{\alpha_{2}}{4D^{2\alpha_{1}}}\frac{1}{\left[1+2\alpha_{1}\right]\alpha_{1}} 
\times 
\\ & \times &  \left[\frac{1}{\left|\frac{1}{D}\right|^{2\alpha_{1}}}
-\frac{1}{\left|\tau\right|^{2\alpha_{1}}}\right], 
\\ 
\delta \chi\left(i\omega_{n}\neq0\right) & = & 
\frac{\alpha_{2}^{2}}{\left[1+2\alpha_{1}\right]2\alpha_{1}D^{2\alpha_{1}}
\left|\omega_{n}\right|^{1-2\alpha_{1}}} \times \\  
&\times  &\left\{ 
  \Gamma\left(1-2\alpha_{1}\right)\sin\left[\pi\alpha_{1}\right]\right\} . 
\end{eqnarray*} 
Expanding for $\alpha_{1} \ll 1$ and $\alpha_{1}\ln(D/|\omega_{n}|) \ll 1$ 
we find: 
\begin{eqnarray} 
\delta \chi\left(i\omega_{n}\neq0\right) & = & 
\frac{\pi\alpha_{1}\alpha_{2}}{\left|\omega_{n}\right|}\left[
\left(\mathcal{C}-1\right)+\ln\left|\frac{\omega_{n}}{D}\right|\right], 
\label{eq:xik1k2O4} 
\end{eqnarray} 
where $\mathcal{C}\approx0.57772$ is the Euler-Gamma constant. The 
susceptibility in fourth order in the coupling constants is  
the sum of Eq.~(\ref{eq:xik1k2O4}) and the Fourier transform of Eq.~(\ref{eq:cork2O4}), 
\begin{widetext} 
\begin{eqnarray} 
\chi\left(i\omega_{n}\right) & = & 
\frac{1}{4}\delta\left(\omega_{n}\right)+\frac{\pi\alpha_{2}}{2}\frac{1}{\left|\omega_{n}\right|} 
\left[1-2\left(\alpha_{2}+\alpha_{1}\right)\ln\left(\frac{D}{\left|\omega_{n}\right|}\right)
+2\mathcal{C}\alpha_{2}+2\left(\mathcal{C}-1\right)\alpha_{1}+ 
{\cal O}[\alpha_2^2,\alpha_1^2]\right] . 
\label{eq:xi-T0} 
\end{eqnarray} 
\end{widetext} 
 
\subsection{Asymptotic regime of $h\to\infty$}\label{sec:RPA}

We represent the spin variables in (\ref{zerofield}) in terms of two spinless fermions: 
\begin{eqnarray*} 
S_{1} & = & \left(a^{\dagger}b+b^{\dagger}a\right)/2\\ 
S_{2} & = & -i\left(a^{\dagger}b-b^{\dagger}a\right)/2\\ 
S_{3} & = & \left(a^{\dagger}a-b^{\dagger}b\right)/2 
\end{eqnarray*} 
and add to the action an imaginary chemical potential, $i\omega_{0}=i \pi T/2.$ 
Working with this formalism we can use Wick's theorem and the standard 
diagrammatic technique. The action is rewritten as 
\begin{eqnarray*} 
S & = & S_{0}\left(\phi_{1},\phi_{2}\right)+\int d\tau\, 
a^{*}\left(\tau\right)\left[\partial_{\tau}-\mathtt{z}\right]a\left(\tau\right)+ 
\\ 
 & + & 
 b^{*}\left(\tau\right)\left[\partial_{\tau}+\mathtt{z}^{*}\right]b\left(\tau\right)+\sqrt{2\pi}\left[\sqrt{\alpha_{1}}\partial_x\phi_{1}\left(0,\tau\right)\right. 
\\ 
 & - & \left.i\sqrt{\alpha_{2}}\partial_{x}\phi_{2}\left(0,\tau\right)\right]a*\left(\tau\right)b\left(\tau\right)+\mathtt{h}.\mathtt{c}.\,,\end{eqnarray*} 
where $\mathtt{z}=i\omega_{0}+\frac{h}{2}$. We define the following propagators, 
\begin{eqnarray*} 
\mathcal{G}_{a}^{\left(0\right)}\left(i\omega _{n}\right) &= &  
  \frac{-1}{i\omega _{n}-\mathtt{z}},\\ 
  \mathcal{G}_{b}^{\left(0\right)}\left(i\omega _{n}\right) & =& 
  \frac{-1}{i\omega _{n}+\mathtt{z}^{*}},\\ 
  \mathcal{D}_{1,2}^{\left(0\right)}\left(i\omega _{n} \neq 0\right) & =& 
  -\frac{|\omega _{n}|}{2 \pi}\arctan{\left(\frac{D}{|\omega _{n}|}\right)} 
  \, ,  
\end{eqnarray*} 
for the fields $a(\tau)$, $b(\tau)$, and the boundary field $\phi_{1,2}(0,\tau)$. 
 
From the propagators we immediately derive the zeroth order part of the susceptibility:  
\begin{eqnarray*}  
\chi_{0}\left(i\omega _{n} \right) & = & 
\frac{T}{4}\sum 
  _{p_{n}}\mathcal{G}_{a}^{\left(0\right)}\left(i\omega 
    _{n}+ip_{n}\right)\mathcal{G}_{b}^{\left(0\right)}\left(ip_{n}\right)\\ 
  &  & +\mathcal{G}_{b}^{\left(0\right)}\left(i\omega 
    _{n}+ip_{n}\right)\mathcal{G}_{a}^{\left(0\right)}\left(ip_{n}\right)\\ 
  & = & \frac{1}{4 }\left(\frac{1}{i\omega _{n}+h}-\frac{1}{i\omega 
      _{n}-h}\right)\\ & = & \frac{1}{2}\frac{h}{\omega 
    _{n}^{2}+h^{2}} . 
\end{eqnarray*} 
 
A simple perturbative calculation will fail to capture the physics and the correct behavior of the 
susceptibility. Following the standard prescription we will sum the infinite 
series of bubble diagrams in the RPA. Let us first consider the second order 
bubble diagrams. From the definition of the propagators and assuming  
$\left|\omega_{n}\right|\ll D$, we obtain:  
\begin{eqnarray*}  
\delta \chi_1\left(i\omega _{n} \neq 0 \right) & = & 
- \frac{\pi \alpha_{1}}{2} 
\frac{h^{2}}{\left(\omega_{n}^{2}+h^{2}\right)^{2}}|\omega_{n}| 
\\& = & -2\pi \alpha_{1}\left[\chi_{0}\left(i\omega _{n} 
  \right)\right]^{2} |\omega _{n}|\\ & &  
\\ \delta \chi_2\left(i\omega 
  _{n}\neq 0 \right) & = &  
\frac{\pi \alpha_{2}}{2} \frac{\omega_{n}^{2}}{\left(\omega_{n}^{2}+h^{2}\right)^{2}}|\omega_{n}| \\ & = & 2\pi \alpha_{2}\left[ \chi_{0} \left( i\omega _{n} \right) \right]^{2} \left( \frac{\omega _{n}}{h}\right)^{2} |\omega _{n}| . 
\end{eqnarray*} 
 
The bubble diagrams in fourth and sixth order can be calculated in a straightforward way: 
\begin{eqnarray*} 
\frac{\chi_{\mathtt{RPA}}^{\left(4\right)}\left(i\omega_{n}\right)}{\left[\chi_0 
 \left(i\omega_{n}\right)\right]^{3}} 
 & = &  
4\pi^{2}\omega_{n}^{2}\left[\alpha_{1}^{2}-\alpha_{2}\left(\frac{\omega_{n}}{h}\right)^{2}\left(2\alpha_{1}+\alpha_{2}\right)\right]\\ 
\frac{\chi_{\mathtt{RPA}}^{\left(6\right)}\left(i\omega_{n}\right)}{ 
\left[\chi_0\left(i\omega_{n}\right)\right]^{4}} 
& = &  
-8\pi^{3}\left|\omega_{n}\right|^{3}\left[\alpha_{1}^{3}-\alpha_{2}\left(\frac{\omega_{n}}{h}\right)^{2}\left(3\alpha_{1}^{2} 
 \right. \right. 
\nonumber 
\\ 
&+& 
 \left. 
 \left. 
 2\alpha_{1}\alpha_{2}+\alpha_{2}^{2}\right)+\alpha_{1}\alpha_{2}^{2}\left(\frac{\omega_{n}}{h}\right)^{4}\right]\, . 
\end{eqnarray*} 
For $h\gg\alpha_{2}\omega_{n}$, we can simplify these results 
and sum the geometric series,  
\begin{eqnarray*} 
\chi_{\mathtt{RPA}}\left(i\omega_{n}\right) & \approx & 
\frac{\chi_0 \left(i\omega_{n}\right)}{1+2\pi\alpha_{1}\left|\omega_{n}\right|\chi^{\left(0\right)}\left(i\omega_{n}\right)}, 
\\ 
 & \approx & 
 \frac{\left(h/2\right)}{h^{2}+\omega_{n}^{2}+\pi h \alpha_{1}\left|\omega_{n}\right|}. 
\end{eqnarray*} 
The zero temperature susceptibility in the RPA approximation (for low 
frequencies and high magnetic fields) is obtained by the analytical continuation, 
\begin{eqnarray*} 
\frac{\chi_{\mathtt{RPA}}^{\prime\prime}\left(\omega\right)}{\omega} & 
\approx & 
\frac{\pi}{2}\frac{\alpha^2_{1}h}{\left(\omega^{2}-h^{2}\right)^{2}+\pi^{2}h\alpha_{1}^{2}\omega^{2}}. 
\end{eqnarray*} 
If we define the decoherence time  
\begin{eqnarray} 
T_2^{-1}=\frac{\pi}{2}h\alpha^2_{1} \, , 
\end{eqnarray} 
we can identify the functional form obtained in Eq.~(\ref{eq:Bloch-sol}), 
\begin{eqnarray*} 
\frac{\chi_{\mathtt{RPA}}^{\prime\prime}\left(\omega\right)}{\omega} & 
\approx & 
\frac{h/T_2}{\left(\omega^{2}-h^{2}\right)^{2}+4 \omega^2/T_2^{2}}, 
\end{eqnarray*} 
that leads to (\ref{rpa}) if one replaces: 
\begin{eqnarray} 
h &\to& T_A 
\nonumber 
\\ 
\alpha_1 &\to& \alpha^* 
\nonumber 
\\ 
T_2^{-1} &\to& \tau^{-1} = \frac{\pi}{2} T_A (\alpha^*)^2\, . 
\end{eqnarray}


\begin{thebibliography}{50}
\bibitem{note_ope}The same equation can be obtained in a straightforward way 
  from the product operator expansion (OPE) for $A_1^{\dagger}$ and $A_1$ 
(see [\onlinecite{zarand}]).
\end{thebibliography}
\end{document}